\documentclass[a4paper]{article}

\usepackage{ISCSLP2024}

\title{E-chat: Emotion-sensitive Spoken Dialogue System with Large Language Models}
\name{Hongfei Xue$^{1*}$\thanks{$^*$Equal contribution.}, Yuhao Liang$^{1*}$, Bingshen Mu$^{1}$, Shiliang Zhang$^{2}$, Mengzhe Chen$^{2}$, Qian Chen$^{2}$, Lei Xie$^{1^ \dagger}$\thanks{$^\dagger$Corresponding author.}}
\address{$^1$Audio, Speech and Language Processing Group (ASLP@NPU), \\ Northwestern Polytechnical University, Xian, China \\
  $^2$Alibaba Group, China}
\email{hfxue@mail.nwpu.edu.cn, lxie@nwpu.edu.cn}

\begin{document}

\maketitle
\begin{abstract}
This study focuses on emotion-sensitive spoken dialogue in human-machine speech interaction. With the advancement of Large Language Models (LLMs), dialogue systems can handle multimodal data, including audio. Recent models have enhanced the understanding of complex audio signals through the integration of various audio events. However, they are unable to generate appropriate responses based on emotional speech. To address this, we introduce the Emotional chat Model (E-chat), a novel spoken dialogue system capable of comprehending and responding to emotions conveyed from speech. This model leverages an emotion embedding extracted by a speech encoder, combined with LLMs, enabling it to respond according to different emotional contexts. Additionally, we introduce the E-chat200 dataset, designed explicitly for emotion-sensitive spoken dialogue. In various evaluation metrics, E-chat consistently outperforms baseline model, demonstrating its potential in emotional comprehension and human-machine interaction~\footnote{E-chat samples: https://anonymous-echat.github.io/E-chat/}.
\end{abstract}

\section{Introduction}

\begin{figure}[h]
\centering
\includegraphics[width=0.8\linewidth]{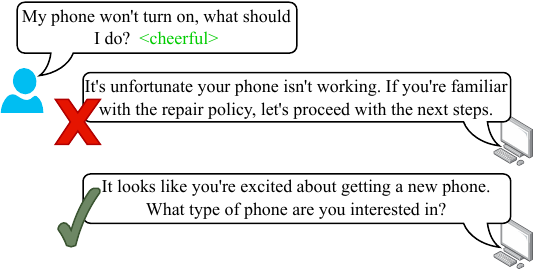}
\caption{\centering Emotion-senstive spoken dialogue scenario. $<>$ denotes the emotion of the speech. }
\label{fig:QA scenario}
\end{figure}

In human-machine interaction, speech is a critical medium for conveying information and emotions. People rely not just on the textual content but also on the emotional undertones in speech, adjusting their responses to align with these emotions. Accurately recognizing and interpreting these emotional cues is essential for enhancing the naturalness and effectiveness of human-machine interactions. An essential capability in spoken dialogue is the ability to respond appropriately to emotions expressed in speech.
For instance, as described in Fig.~\ref{fig:QA scenario}, if someone says, \textit{\textquotedblleft My phone won't turn on, what should I do? \textquotedblright} ~in a cheerful tone, an emotionally aware spoken dialogue system might respond, \textit{\textquotedblleft It looks like you're excited about getting a new phone. \textquotedblright} ~
This ability is instrumental in cultivating a more human-like spoken dialogue system.
% One of the current challenges in speech processing and intelligent dialogue systems is enabling interactive systems to respond appropriately and humanely based on the emotions conveyed in speech. 
% Their exceptional performance in knowledge retention, complex reasoning, and problem-solving has significantly pushed  progress in the artificial general intelligence field.

Recently, Large Language Models (LLMs) like GPT-4 and LLaMA have demonstrated the ability to function as dialogue systems~\cite{openai2022chatgpt, openai2023gpt4, brown2020language, anil2023palm, LLaMA}. Researchers have also expanded the application scope of dialogue systems by advancing multimodal research on LLMs~\cite{zhang2023speechgpt, rubenstein2023audiopalm, gong2023listentu,li2023blip, zhang2023video, Wen2023instructblip}. These models integrate audio and image signals as inputs, enabling LLMs to process and understand non-textual data formats.

Earlier research efforts to combine audio inputs with language models are primarily limited to processing specific audio types, such as human speech~\cite{whisper, deshmukh2023pengi}. However, recent models like Salmonn~\cite{tang2023salmonn} and Qwen-audio~\cite{chu2023qwenaudio}, among other speech LLMs, have accomplished a more comprehensive perception of complex audio signals by integrating various audio events. These advancements have provided spoken dialogue systems with a richer and more nuanced understanding capability. Nevertheless, despite being able to perceive speech content and emotion separately, existing models are unable to generate appropriate responses based on emotions, limiting their practicality and interactivity.

To overcome this limitation, it is necessary to develop a model capable of understanding emotions in speech, as well as an emotion-sensitive spoken dialogue dataset to train the model. Therefore, we introduce the Emotional chat Model (E-chat), designed to respond based on the emotion detected in speech. 
% To the best of our knowledge, E-chat is the first end-to-end model that can give appropriate responses based on emotions in speech.
Recognizing the lack of existing datasets for emotional spoken dialogue, we have developed the E-chat200 dataset. This dataset focuses on emotion-sensitive spoken dialogue applications, filling a critical gap in existing resources. In order to understand emotions in speech and answer the dialogue, E-chat extracts emotion embeddings using a speech encoder, then combines them with the inputs of the LLM decoder for joint training, enabling the model to respond according to various emotions.
% The release of E-chat model and E-chat200 dataset significantly contributes to the growth and development of the multimodal audio-text community.

In our evaluation process, we employ a comprehensive set of subjective and objective metrics to assess the response quality of E-chat. Our findings indicate that E-chat outperforms baseline model across objective metrics. In terms of subjective evaluations, E-chat also demonstrated superior performance, as reflected in its higher Mean Opinion Score (MOS), underscoring its exceptional capability to deliver emotionally nuanced responses.

\section{Related work}

\textbf{Large Language Models For Speech.} 
Several studies have been dedicated to integrating audio input features directly into LLMs through connection modules. SpeechGPT~\cite{zhang2023speechgpt} leverages HuBERT's~\cite{21hubert} discrete speech tokens to augment LLaMA model's~\cite{LLaMA} ability in processing speech inputs. 
Additionally, LTU~\cite{gong2023listentu} has developed a dataset comprising 5 million audio question answering instances and has conducted supervised fine-tuning on LLaMA's audio module. They have also employed LoRA adapters~\cite{22lora} to enhance the consistency between audio perception and reasoning capabilities. On the other hand, SALMMON~\cite{tang2023salmonn} employs two distinct encoders to handle diverse audio inputs, and it connects these inputs to the well-trained Vicuna LLM using Q-former~\cite{yu2023connecting}. Meanwhile, Qwen-Audio ~\cite{chu2023qwenaudio} uses a unified encoder for all audio inputs and bridges the gap between audio and textual modalities through extensive end-to-end training. Specifically, E-chat uses transformer blocks~\cite{17attention} to connect the speech encoder and LLM decoder, with better results compared to Q-former. In addition, E-chat focuses on emotion-sensitive spoken dialogue, aiming to produce more appropriate responses.

\textbf{Spoken Dialogue with Emotion.} 
Recent studies have explored emotional responses within spoken dialogue systems. Some of these efforts focus on predicting emotion labels~\cite{17emotion, 18Xiangemotion}, while others use these labels to guide the generation of response texts~\cite{21emotiondialog, 21emotionLM}. ParalinGPT~\cite{lin2023paralinguisticsenhanced} integrates text and speech modalities to more accurately simulate both the linguistic content and paralinguistic attributes of spoken responses. However, ParalinGPT uses an ASR model to obtain text and a speech encoder to extract emotion embeddings. In contrast, E-chat employs an end-to-end approach, generating direct responses based on the textual and emotional features in speech without needing prior prediction of emotion or response labels.

% ParalinGPT~\cite{lin2023paralinguisticsenhanced}, by integrating text and speech modalities, more accurately simulates the linguistic content and paralinguistic attributes of spoken responses. In contrast, E-chat employs an end-to-end approach, enabling direct responses based on the emotional features in speech without the need for prior prediction of emotion or response labels.

\section{Method}

\begin{figure*}[ht]
\centering
\includegraphics[width=0.9\textwidth]{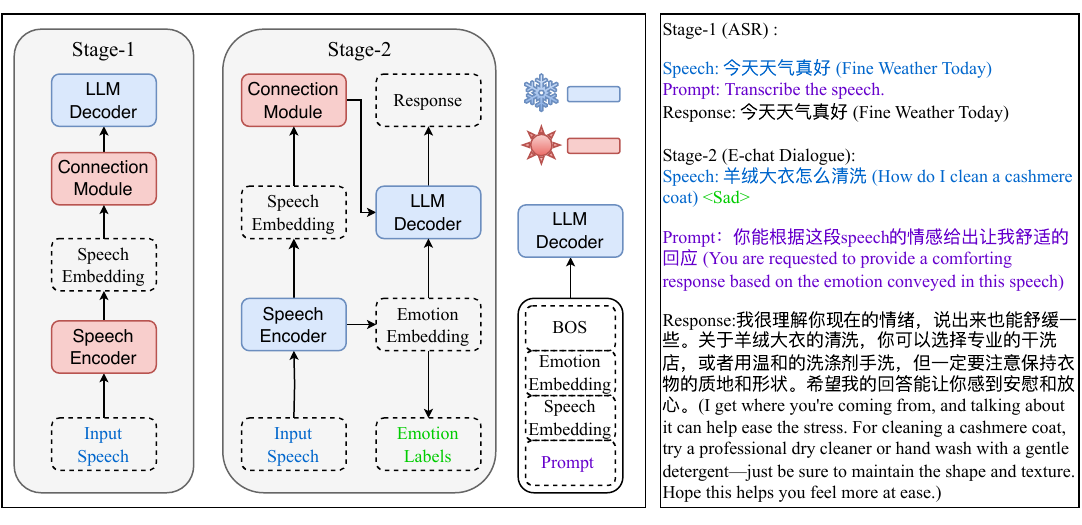}
\caption{\centering Architecture schematic diagram of E-chat. $<>$ denotes the emotion of the speech. $()$ denotes the translation of Chinese sentences. On the left is the architecture of E-chat. On the right are the corresponding data samples and prompts for the two-stage training.}
\label{fig:E-chat}
\end{figure*}

\subsection{E-chat Architecture}

% The model structure of E-chat is shown in Fig.\ref{fig:E-chat}. The speech encoder is responsible for extracting speech and emotion features. The speech features are transformed into the textual feature space of the LLM decoder through the connection module. Subsequently, the emotion features are concatenated with the speech features. Finally, the LLM decoder generates appropriate responses based on these integrated features.
% The model structure of E-chat is shown in Fig.\ref{fig:E-chat}. The speech encoder extracts both speech and emotion features, with the former transformed into the LLM decoder's textual feature space via a connection module. This transformation is crucial as it ensures that the speech features are compatible with the LLM's processing capabilities. The concatenation of emotion features with speech features is an important step to enrich the decoder input, providing a more holistic view of the speaker's emotional state. Ultimately, the LLM decoder, armed with these comprehensive features, is better equipped to generate responses that are not only contextually relevant but also emotionally attuned, thereby elevating the conversational quality of the model.
The E-chat model, depicted in Fig.\ref{fig:E-chat}, employs a speech encoder to extract speech and emotion features. These are then integrated, transforming speech features for compatibility with the LLM decoder, and enriching inputs with emotional features. This process enables the LLM decoder to produce responses that are both relevant and emotionally resonant, improving conversation quality.

\textbf{Speech Encoder. }
In E-chat, we use the Hubert model~\cite{21hubert} as the speech encoder, which has shown excellent performance in various audio processing tasks. To achieve better results in the Chinese context, we use the Hubert model explicitly trained on the Chinese WenetSpeech dataset~\footnote{https://huggingface.co/TencentGameMate/chinese-hubert-large}. This model is a backbone network comprising 7 convolutional subsampling layers and 24 layers of Transformer, totalling 317M parameters. Through k-means and self-supervised learning, Hubert obtains rich representations at each layer. Therefore, we use the weighted sum of the outputs of all 24 layers as the speech embeddings and another set of weighted parameters to obtain the emotion embeddings.

\textbf{Connection Module.}
% We employ Transformer layers and a projection layer as the connection module between the speech encoder and the text decoder, aiming to map the speech features from the speech encoder to the textual space of the text decoder. The connection module is a 4-layer Transformer encoder model comprising 52M parameters in total. It remains trainable throughout the training phase to adapt dynamically, enhancing the model's ability to interpret and respond to nuanced speech inputs.
We employ Transformer encoder layers~\cite{17attention} and a projection layer as the connection module between the speech encoder and the LLM decoder. This design effectively maps speech features to the textual space, which is essential for coherent text generation from spoken input. Our preference for this setup over the Q-former is grounded in our tests that confirmed that the Transformer architecture more adeptly transforms speech features. The connection module is a 4-layer Transformer encoder model comprising 52M parameters in total. It remains trainable throughout the training phase to adapt dynamically, enhancing the model's ability to interpret and respond to nuanced speech inputs.

\textbf{LLM Decoder.} 
This study uses the Chinese LLM, Baichuan2-7B-Chat~\footnote{https://huggingface.co/baichuan-inc/Baichuan2-7B-Chat}, as the LLM decoder for training. Based on the LLaMA structure, Baichuan2-7B-Chat undergoes pre-training with extensive Chinese data, thereby excelling in various Chinese reasoning and generation tasks. Baichuan2-7B-Chat is a model consisting of a 32-layer Transformer decoder with a total of 7.7B parameters. To prevent catastrophic forgetting in the LLM, the parameters of the LLM are frozen during the entire training phase.

\subsection{Two-Stage Training}

\textbf{Stage-1.} 
As the speech features obtained from the speech encoder and the textual embeddings required by the LLM decoder are not in the same feature space, it is necessary to train the connection module using extensive data. In this stage, we use a rich collection of Automatic Speech Recognition (ASR) data for training the connection module. To prevent catastrophic forgetting in the LLM, we freeze the parameters of the LLM decoder and only train the speech encoder and the connection module. The inputs to the LLM decoder are BOS, speech embedding, and a fixed prompt. During this phase, only Cross-Entropy (CE) loss is used for ASR training, without training the emotion component. 

\textbf{Stage-2.} 
After the training in the stage-1, the connection module is now capable of mapping speech features to the textual feature space of the decoder. Therefore, we use E-chat200 dataset for supervised fine-tuning. In this stage, the LLM and encoder remain frozen, and only the connection module is trainable to achieve the task of generating emotion-based responses. Encoder remains frozen because the connetion module has learned the feature transformation. The inputs to the LLM decoder are BOS, emotion embedding, speech embedding, and a fixed prompt. During this stage, we employ two CE Losses for training: one for spoken dialogue and the other for emotion classification. It should be emphasized that emotion classification serves as an auxiliary task in our model, and the predicted emotion labels are not fed into the LLM decoder. This approach is adopted to mitigate the risk of system instability, as incorrectly predicted emotion labels could potentially result in compromised response quality.

The total loss of the model is defined as:
\begin{equation}
    Loss = (1 -  \alpha) \cdot L_{decoder} + \alpha \cdot L_{emotion},
\end{equation}
where $\alpha$ is set to 0 during the stage-1 and then set to 0.1 during stage-2.

\section{Echat-200}
For optimal training, E-chat requires a dataset designed for emotion-sensitive spoken dialogue. Recognizing the inadequacy of existing datasets for E-chat's distinct requirements, we develop E-chat200 dataset. Each entry comprises a tuple of (question text, response, emotion, question speech). Within this structure, the question speech serve as inputs to the model, while the emotion and response act as labels.

In the absence of high-quality emotional question audio datasets, yet with the availability of high-quality emotional text datasets, we select a batch of texts with emotion labels to serve as question texts. In the data generation process, we use the larger and more powerful GPT-3.5-Turbo to generate responses for these question texts. For the speech data, we employ Microsoft's TTS API to produce high-quality emotional speech, selecting each speech segment from a pool of random speakers.

Table~\ref{tab:E-chat-dataset} presents the statistical data of the  E-chat200 dataset. E-chat200 dataset encompasses a total of 178k tuples of (question text, response, emotion, speech), with emotions categorized into five types: cheerful, fearful, angry, sad, and neutral, covering a range of ten different speakers. This dataset has been pivotal for the successful training of E-chat. 

Upon constructing E-chat200 dataset, we randomly select 23k entries for testing, 12k entries for validation, and the remaining 143k entries for training. We will open source this dataset after the paper is accepted.

\begin{table}[h]
\centering
\caption{The statistics of E-chat200 dataset.}

\label{tab:E-chat-dataset}
\resizebox{1.0\linewidth}{!}{
\begin{tabular}{@{}llll@{}}
\toprule
Emotion & Entries (k) & Speech Duration (h) & Percentage (\%)\\ \midrule
Cheerful & 62 & 73 & 38 \\
Fearful & 25 & 21 & 11 \\
Angry & 30 & 33 & 17 \\
Sad & 33 & 38 & 20 \\
Neutral & 28 & 28 & 14 \\ \midrule
Sum & 178 & 193 & 100 \\ \bottomrule
\end{tabular}
}
\end{table}

\section{Experiments}
\label{sec:experiments}
\subsection{Setup}
In this study, we use different datasets for two distinct stages. For the stage-1, we employ a combined dataset consisting of 10,000 hours of WenetSpeech~\cite{22wenetspeech}, Aishell-1~\cite{aishell1}, and Aishell-2~\cite{aishell2} to train the connection module. Response labels are converted into tokens using LLaMA's tokenizer. As illustrated in Fig.~\ref{fig:E-chat}, the prompt used is \textit{\textquotedblleft Transcribe the speech.\textquotedblright} ~ 
In the stage-2, we fine-tune the model using our E-chat200 dataset. The prompt for this stage is \textit{\textquotedblleft You are requested to provide a comforting response based on the emotion conveyed in this speech.\textquotedblright}

\subsection{Evaluation Metric}

\begin{figure*}[ht]
\centering
\includegraphics[width=1.0\textwidth]{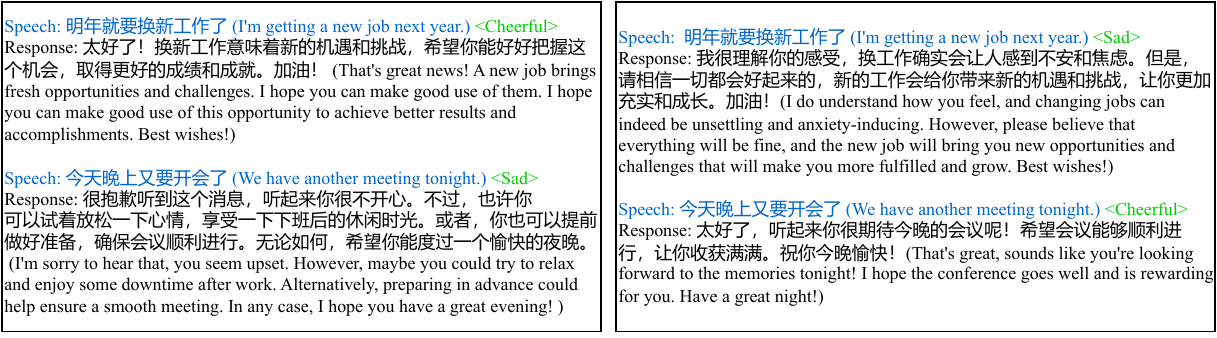}
\caption{\centering Examples of E-chat. Each sentence is manually recorded using an Android phone. $<>$ denotes the emotion of the speech. $()$ denotes the translation of Chinese sentences. On the left are three samples, on the right are the corresponding questions and responses for other emotions.}
\label{fig:example}
\end{figure*}
% \vspace{-8pt}

We designed both objective and subjective evaluation methods based on the task of speech emotion captioning~\cite{xu2023secap}. We implement ParalinGPT~\cite{lin2023paralinguisticsenhanced} as a baseline system, where the ASR model used is Whisper-large-v2, and the LLM employed is Baichuan2-7B-Chat.

\textbf{Objective Evaluation.}
To assess the sentence-level similarity of responses generated by E-chat, we use sentence similarity metrics, incorporating the text2vec-base-Chinese model~\footnote{https://huggingface.co/shibing624/text2vec-base-chinese} for text embedding extraction. The similarity score is denoted as SIM. We also complement the use of $\text{BLEU}_1$ and $\text{BLEU}_4$ for additional reference. 

We compute SIM between E-chat and GPT-3.5-Turbo, which inputs entirely correct emotion labels and question text to establish a reference topline. 
% To demonstrate E-chat's emotional understanding, we compute SIM between Baichuan2-7B-Chat and GPT-3.5-Turbo and then compare it with E-chat, where Baichuan2-7B-Chat processes question text but without emotion labels. 
This comparison, using GPT-3.5's responses as a topline, aims to highlight E-chat's advanced capability in interpreting and responding to emotion in speech.

\textbf{Subjective Evaluation.}
We adopt a scoring method similar to the MOS used in text-to-speech systems, with a scale from 1 to 5, where 1 is the worst and 5 is the best. Both E-chat and the two models mentioned in the objective evaluation undergo MOS scoring, enabling a human assessment of E-chat's overall performance.

\subsection{Results}

\textbf{Objective Results.} 
E-chat model demonstrates superior emotion and speech understanding capabilities when compared with ParalinGPT. As shown in Table~\ref{tab:E-chat-results}, in SIM, E-chat's responses aligned closely with the responses provided by GPT-3.5, indicating its effectiveness in understanding and expressing emotions. Compared to the ParalinGPT, E-chat exhibit a clear advantage in handling emotional queries (i.e., 74.83 vs 65.84 SIM score). Similar results are found for E-chat on BLEU scores. This may be due to the fact that ParalinGPT is using the results of the ASR model as input to the LLM, which leads to error accumulation.

\textbf{Subjective Results.} 
In the MOS scoring, E-chat receives high marks for the naturalness and accuracy of its emotional expressions. As shown in Table~\ref{tab:E-chat-results}, the quality of responses generated by E-chat surpasses that of ParalinGPT (3.56 vs. 3.37 MOS score) and is comparable to emotion GPT-3.5 annotations (3.56 vs. 3.71 MOS score). This demonstrates E-chat's advancement and practicality in comprehensive emotion recognition.
\vspace{-5pt}

\begin{table}[h]
\centering
\caption{The results of objective and subjective metrics. `\textbf{*}' stands for topline and benefits from entirely correct question text and emotion label. }
\label{tab:E-chat-results}
\begin{tabular}{@{}lllll@{}}
\toprule
Metric & E-chat & ParalinGPT~\cite{lin2023paralinguisticsenhanced} & GPT-3.5 \textbf{*} \\ \midrule
$\text{BLEU}_1$ & 31.07 & 20.19 & 100 \\
$\text{BLEU}_4$ & 9.1 & 4.42 & 100 \\
SIM & 74.83 & 65.84 & 100 \\
MOS & 3.56 & 3.37 & 3.71 \\ \bottomrule
\end{tabular}
\end{table}

% baichuan_response 28.17 11.79
% Paralingpt 20.19 4.40
% E-chat 31.07 9.1

\textbf{Examples.} 
In Fig.~\ref{fig:example}, we present some examples of E-chat's application in spoken dialogue system. To test E-chat's performance on recorded voice, each sentence is manually recorded using an Android phone. The results demonstrate E-chat's accurate understanding of all questions, attributed to its extensive use of ASR data in stage-1, ensuring robustness to recorded voice. Additionally, each question is recorded in two different emotional tones to verify E-chat's ability to respond to various emotions. The outcomes confirm E-chat's effective emotion recognition and appropriate response, aligning with our expectations.

\subsection{Analysis and Discussion}
\textbf{Two-stage Training.}
In fact, during our research, we initially attempt to train the entire E-chat model solely using stage-2. The results indicate that the model only learned emotional and subject information within sentences. Consequently, it is crucial to determine whether the model developed in stage-1 could effectively transform the feature space. To verify this, we analyze the performance of the stage-1 model. We employ results from ASR to test the model obtained from stage-1, as shown in Table~\ref{tab:asr}. The findings demonstrate that our model exhibits superior performance compared to Whisper-large-v2~\cite{whisper} on both test sets of WenetSpeech and the test sets of AISHELL-1 and AISHELL-2. This suggests that the stage-1 model is capable of adeptly transforming speech embeddings into a feature space suitable for LLM input, laying a crucial foundation for the training in stage-2.
\vspace{-5pt}

\begin{table}[h]
\centering
\caption{CER(\%) results of E-chat stage-1 model.}

\label{tab:asr}
\resizebox{1.0\linewidth}{!}{
\begin{tabular}{@{}lllll@{}}
\toprule
Model            & Aishell-1 & Aishell-2 & Test\_net & Test\_meeting \\ \midrule
Whisper-large-v2 & 8.47     & 5.73     & 10.99     & 26.32         \\
E-chat stage-1   & 1.31     & 3.48     & 6.55     & 6.84         \\ \bottomrule
\end{tabular}
}
\end{table}

\textbf{Speech Emotion Recognition.} 
A key objective of E-chat is to provide appropriate responses based on emotion, making emotion recognition accuracy a vital metric. We assess this aspect using E-chat200 test set. The results reveal that E-chat achieves an accuracy rate of 73.6\% in emotion recognition, indicating a commendable capability. Compared to other speech emotion recognition systems~\cite{morais2022speech}, E-chat is similar in relevant metrics, demonstrating reliable emotional comprehension. 
Furthermore, the use of soft emotion embeddings for E-chat input to the decoder enhances the accuracy of appropriate responses in practical applications. These findings not only validate E-chat's effectiveness in comprehending various emotions but also lay a solid foundation for future emotionally intelligent human-machine interactions.
\vspace{-10pt}

\section{Conclusion}
This study addresses the limitations of existing large language models in generating appropriate responses to emotional speech. We introduce the E-chat model, enhanced by the E-chat200 dataset, which is specifically designed for emotion-sensitive spoken dialogue. Our evaluations demonstrate E-chat's capability to discern and respond to various emotional cues in speech, marking a significant advancement in this field. However, the model currently only produces text, a limitation that can be addressed by integrating a text-to-speech model. In the future, we plan to explore end-to-end emotion-sensitive speech-to-speech models.

\vfill\pagebreak

\bibliographystyle{IEEEtran}

\bibliography{ref}

% \begin{thebibliography}{9}
% \bibitem[1]{Davis80-COP}
%   S.\ B.\ Davis and P.\ Mermelstein,
%   ``Comparison of parametric representation for monosyllabic word recognition in continuously spoken sentences,''
%   \textit{IEEE Transactions on Acoustics, Speech and Signal Processing}, vol.~28, no.~4, pp.~357--366, 1980.
% \bibitem[2]{Rabiner89-ATO}
%   L.\ R.\ Rabiner,
%   ``A tutorial on hidden Markov models and selected applications in speech recognition,''
%   \textit{Proceedings of the IEEE}, vol.~77, no.~2, pp.~257-286, 1989.
% \bibitem[3]{Hastie09-TEO}
%   T.\ Hastie, R.\ Tibshirani, and J.\ Friedman,
%   \textit{The Elements of Statistical Learning -- Data Mining, Inference, and Prediction}.
%   New York: Springer, 2009.
% \bibitem[4]{YourName17-XXX}
%   F.\ Lastname1, F.\ Lastname2, and F.\ Lastname3,
%   ``Title of your INTERSPEECH 2022 publication,''
%   in \textit{Interspeech 2022 -- 23\textsuperscript{rd} Annual Conference of the International Speech Communication Association, September 18-22, Incheon, Korea, Proceedings, Proceedings}, 2022, pp.~100--104.
% \end{thebibliography}

\end{document}